# Distributed Computing Economics

Jim Gray

Microsoft Research

Microsoft Corporation

One Microsoft Way

Redmond, WA  98052

# Distributed Computing Economics

*Jim Gray,*
*Microsoft Research, San Francisco, California, USA*
*March 2003*

Computing economics are changing. Today there is rough price parity between (1) one database access, (2) ten bytes of network traffic, (3) 100,000 instructions, (4) 10 bytes of disk storage, and (5) a megabyte of disk bandwidth. This has implications for how one structures Internet-scale distributed computing: one puts computing as close to the data as possible in order to avoid expensive network traffic.

## The Cost of Computing

**Computing is Free**. The world's most powerful computer is free (SETI@Home is a 54 teraflops machine).[1] Google freely provides a trillion searches per year to the world's largest online database (2 petabytes). Hotmail freely carries a trillion eMail messages per year. Amazon.com offers a free book search tool. Many sites offer free news and other free content. Movies, sports events, concerts, and entertainment are freely available via television.

**Actually, it's not free, but most computing is now so inexpensive that advertising can pay for it**. The content is not really free; it is paid for by advertising. Advertisers routinely pay more than a dollar per thousand impressions (CPM). If Google or Hotmail can collect a dollar per CPM, the resulting billion dollars per year will more than pay for their development and operating expenses. If they can deliver a search or a mail message for a few micro-dollars, the advertising pays them a few milli-dollars for the incidental "eyeballs". So, these services are not free – advertising pays for them.

**Computing costs hundreds of billions of dollars per year:** IBM, HP, Dell, Unisys, NEC, and Sun each sell billions of dollars of computers each year. Software companies like Microsoft, IBM, Oracle, and Computer Associates sell billions of dollars of software per year. **So, computing is obviously not free**.

**Total Cost of Ownership (TCO) is more than a trillion dollars per year. Operations costs far exceed capital costs.** Hardware and software are minor parts of the total cost of ownership. Hardware comprises less than half the total cost; some claim less than 10% of the cost of a computing service. So, the real cost of computing is measured in trillions of dollars per year.

**Megaservices like Yahoo!, Google, and Hotmail have relatively low operations staff costs.** These megaservices have discovered ways to deliver content for less that the milli-dollar that advertising will fund. For example, in 2002 Google had an operations staff of 25 who managed its two petabyte ($2^{15}$ bytes) database and 10,000 servers spread across several sites. Hotmail and Yahoo! cite similar numbers – small staffs manage ~300 TB of storage and more than ten thousand servers.

**Most applications do not benefit from megaservice economies of scale**. Other companies report that they need an administrator per terabyte, an administrator per 100 servers, and an administrator per gigabit of network bandwidth. That would imply an operations staff of more than two thousand people to operate Google – nearly ten times the size of the company.

**Outsourcing is seen as a way for smaller services to benefit from megaservice efficiencies**. The outsourcing business evolved from service bureaus through timesharing and is now having a renaissance. The premise is that an outsourcing megaservice can offer routine services much more efficiently than an in-house service. Today, companies routinely outsource applications like payroll, insurance, web presence, and eMail.

---

1 This paper makes broad statements about the economics of computing. The numbers are fluid – (costs change every day.) They are approximate to within factor of 3. For this specific fact: SETI@Home averaged 54 teraflops (floating point operations per second) on (1/26/2003) handily beating the sum of the combined peak performance of the top four of the TOP500 supercomputers registered at http://www.top500.org/ on that day.

**Outsourcing has often proved to be a shell game –moving costs from one place to another.**
Loudcloud and Exodus trumpeted the benefits of outsourcing. Now Exodus is bankrupt and Loudcloud is gone. Neither company had a significant competitive advantage over in-house operations. Outsourcing works when it is a service business where computing is central to operating an application and supporting the customer – a high-tech low-touch business. It is difficult to achieve economies-of-scale unless the application is nearly identical across most companies – like payroll or email. Some companies, notably IBM, Salesforce.com, Oracle.com and others are touting outsourcing, *On Demand Computing*, as an innovative way to reduce costs. There are some successes, but many more failures. So far there are few outsourced megaservices – payroll and eMail are the exception rather than the rule.

**SETI@Home sidesteps operations costs and is not funded by advertising**. SETI@Home is a novel kind of outsourcing. It harvests some of the free (unused) computing available in the world. SETI@Home "pays" for computing by providing a screen saver, by appealing to people's interest in finding extra-terrestrial intelligence, and by creating competition among teams that want to demonstrate the performance of their systems. This currency bought 1.3 *million* years of computing; it bought 1.3 thousand years of computing on 3$^{rd}$ February 2003. Indeed, some SETI@Home results have been auctioned at eBay. Others are emulating this model for their compute-centric applications (e.g. Protein@Home or ZetaGrid.net).

**Grid computing hopes to harvest and share Internet resources**. Most computers are idle most of the time, disks are ½ full on average, and most network links are under-utilized. Like the SETI@Home model, Grid computing seeks to harness and share these idle resources by providing an infrastructure that allows idle resources to participate in Internet-scale computations [1].

## Web Services

**Microsoft and IBM tout web services as a new computing model – Internet-scale distributed computing**. They observe that the HTTP Internet is designed for people interacting with computers. Traffic on the future Internet will be dominated by computer-to-computer interactions. Building Internet-scale distributed computations requires many things, but at its core it requires a common object model augmented with a naming and security model. Other services can be layered atop these core services. Web services are the evolution of the RPC, DCE, DCOM, CORBA, RMI, … standards of the 1990's. The main innovation is an XML base that facilitates interoperability among implementations.

**Neither grid computing nor web services have an outsourcing or advertising business model.**
Both are plumbing that enable companies to build applications. Both are designed for computer-to-computer interactions and so have no advertising model – because there are no eyeballs involved in the interactions. It is up the companies to invent business models that make this plumbing useful.

**Web services reduce the costs of publishing and receiving information**. Today, many services offer information as HTML pages on the Internet. This is convenient for people, but programs must resort to screen-scraping to extract the information from the display. If an application wants to send information to another application, it is very convenient to have an information structuring model – an object model, that allows the sender to point to an object (an array, a structure, or a more complex class), and simply send it. The object then "appears" in the address space of the destination application. All the gunk of packaging (serializing) the object, transporting it, and then unpacking it is hidden from sender and receiver. Web services provide this *send-and-object – get-an-object* model. These tools dramatically reduce the programming and management costs of publishing and receiving information.

**So web service is an enabling technology to reduce data interchange costs.** Electronic Data Interchange (EDI) services have been built from the very primitive base of ASN.1. With XML and web services EDI message formats and protocols can be defined in much more concise languages like XML, C#, or Java. Once defined, these interfaces are automatically implemented on all platforms. This dramatically reduces transaction costs. Service providers like Google, Inktomi, Yahoo!, and Hotmail can provide a web service interface that others can integrate or aggregate into personalized digital dashboard and earn revenue from this very convenient and inexpensive service. Many organizations want to publish their information. The World Wide Telescope I have been working on is a small example [2]. These examples are repeated in biology, the social sciences, and the arts. Web services and intelligent user tools are a big advance over publishing a file with no schema (e.g., using FTP).

## Application Economics

**Grid computing and computing-on-demand enable applications that are mobile and that can be provisioned on demand.** What tasks are mobile and can be dynamically provisioned? Any purely computation task is mobile if it is written in a portable language and uses only portable interfaces -- *write once run anywhere* (WORA). Cobol and Java promises WORA. Cobol and Java users can attest that WORA is difficult to achieve, but for the purposes of this discussion, let's assume that it is a solved problem. Then, the question is:

*What are the economic issues of moving a task from one computer to another or from one place to another?*

A task has four characteristic demands:

- **Networking** – delivering questions and answers,
- **Computation** – transforming information to produce new information,
- **Database Access** – access to reference information needed by the computation,
- **Database Storage** – long term storage of information (needed for later access).

**The ratios among these quantities and their relative costs are pivotal.** It is fine to send a GB over the network if it saves years of computation – but it is not economic to send a kilobyte question if the answer could be computed locally in a second.

**To make the economics tangible, take the following baseline hardware parameters [2]:**

| | | |
|---|---|---:|
| 2 | GHz cpu with 2GB ram (cabinet and networking) | $2,000 |
| 200 | GB disk with 100 accesses/ second and 50MB/s transfer | $200 |
| 1 | Mbps WAN link | $100/month |

**From this we conclude that one dollar equates to**

| | | |
|---|---|---|
| = | 1 | $ |
| ~ | 1 | GB sent over the WAN |
| ~ | 10 | Tops tera-cpu instructions |
| ~ | 8 | hours of cpu time |
| ~ | 1 | GB disk space |
| ~ | 10 | M database accesses |
| ~ | 10 | TB of disk bandwidth |

**The ideal mobile task is stateless (no database or database access), has a tiny network input and output, and has huge computational demand.** For example, a cryptographic search problem: given the encrypted text, the clear text, and a key search range. This kind of problem has a few kilobytes input and output, is stateless, and can compute for days. Computing zeros of the zeta function is a good example [3]. Monte Carlo simulation for portfolio risk analysis is another good example. And of course, SETI@Home is a good example: it computes for 12 hours on half a megabyte of input.

**Using parameters above, SETI@Home performed a multi-billion dollar computation for a million dollars** – a very good deal! SETI@Home harvested more than a million cpu years worth a more than a billion dollars. It sent out a billion jobs of ½ MB each. This petabyte of network bandwidth cost about a million dollars. The SETI@Home peers donated a billion dollars of "free" cpu time and also donated $10^{12}$ watt-hours which is about 100M$ of electricity. The key property of SETI@Home is that the ComputeCost:NetworkCost ratio is 10,000:1. It is very cpu-intensive.

---

[2] The hardware prices are typical of web prices, the WAN price is typical of rates paid by large (many Gbps/month) Internet service providers.

**Most web and data processing applications are network or state intensive and are not economically viable as mobile applications**. An FTP server, an HTML web server, a mail server, and Online Transaction Processing (OLTP) server represent a spectrum of services with increasing database state and data access. A 100MB FTP task costs 10 cents and that cost is 99% network cost. An HTML web access costs 10 microdollars and is 88% network cost. A Hotmail transaction costs 10 microdollars and is more cpu intensive so that networking and cpu are approximately balanced. None of these applications fits the cpu-intensive stateless requirement.

**Data loading and data scanning are cpu-intensive; but they are also data intensive, and therefore not economically viable as mobile applications.** Some applications related to database systems are quite cpu intensive: for example data loading takes about 1,000 instructions per byte. The "vision" component of the Sloan Digital Sky Survey that detects stars and galaxies and builds the astronomy catalogs from the pixels is about 10,000 instructions per byte. So, they are break-even candidates: 10,000 instructions per byte is the break-even point according to the economic model above (10 Tops of computing and 1 GB of networking both cost a dollar). It seems the computation should be at 30,000 instructions per byte (a 3:1 cost benefit ratio) before the outsourcing model becomes really attractive. **The break-even point is 10,000 instructions per byte of network traffic or about a minute of computation per MB of network traffic.**

**Few computations exceed that threshold; most are better matched to a Beowulf cluster.** Computational Fluid Dynamics (CFD) is very cpu intensive, but again, CFD generates a continuous and voluminous output stream. To give an example of an adaptive mesh simulation, the Cornell Theory Center has a Beowulf-class MPI job that simulates crack propagation in a mechanical object [4]. It has about 100MB of input, 10GB of output, and runs for more than 7 cpu-years. The computation operates at over one million instructions per byte, and so good is a candidate for export to the WAN computational grid. But, the computation's bisection bandwidth requires that it be executed in a tightly connected cluster. These applications require inexpensive bandwidth available to a Beowulf cluster [5]. In a Beowulf cluster networking is ten thousand times less expensive – which makes it seem nearly free by comparison to WAN networking costs.

**Still, there are some computationally intensive jobs that can use Grid computing.** Render-farms for making animated movies seem to be a good candidate for Grid computing. Rendering a frame can take many cpu hours, so a Grid-scale render farm begins to make sense. For example, Pixar's *Toy Story 2* images are very cpu intensive – a 200 MB image can take several cpu hours to render. The instruction density was 200k to 600k instructions per byte [6]. This could be structured as a grid computation – sending a 50MB task to a server that computes for ten hours and returns a 200MB image.

**BLAST, FASTA, and Smith-Waterson are an interesting case in point – they are mobile in the rare case of a 40 cpu-day computation.** These computations match a DNA sequence against a database like GenBank or SwissProt. The databases are about 50GB today. The algorithms are quite cpu intensive, but they scan large parts of the database. Servers typically store the database in RAM. BLAST is a heuristic that is ten times faster than Smith-Waterson which gives exact results [7, 8]. Most BLAST computations can run in a few minutes of cpu time, but there are computations that can take 720 cpu hours on BLAST and 7200 hours on Smith Waterson. So, it would be economical to send SwisProt (40GB) to a server if it were to perform a 7720 hour computation for free. Typically, it does not make sense to provision a SwissProt database on demand: rather it makes sense to set up dedicated servers (much like Google) that use inexpensive processors and memory to provide such searches. A commodity 40GB SMP server would cost less than $20,000 and could deliver a complex one cpu-hour search for less than a dollar – the typical one minute search would be a few millidollars.

## Conclusions

**Put the computation near the data**. The recurrent theme of this analysis is that "On Demand" computing is only economical for very cpu-intensive (100,000 instructions per byte or a cpu-day-per gigabyte of network traffic) applications.

**How do you combine data from multiple sites?** Many applications need to integrate data from multiple sites into a combined answer. The arguments above suggest that one should push as much of the processing to the data sources as possible in order to filter the data early (database query optimizers call this "pushing predicates down the query tree"). There are many techniques for doing this, but fundamentally it dovetails with the notion that each data source is a web service with a high-level object-oriented interface.

## Caveats

**Beowulf clusters have completely different networking economics**. Render farms, materials simulation, and CFD fit beautifully on Beowulf clusters because there the cost of networking is very inexpensive: a GBps Ethernet fabric costs about 200$/port and delivers 50MBps, so Beowulf networking costs are comparable to disk bandwidth costs – 10,000 times less than the price of Internet transports. That is why rendering farms and BLAST search engines are routinely built using Beowulf clusters. Beowulf clusters should not be confused with Internet-scale Grid computations.

**If telecom prices drop faster than Moore's law, the analysis fails. If telecom prices drop slower than Moore's law, the analysis becomes stronger**. Most of the argument in this paper pivots on the relatively high price of telecommunications. Over the last 40 years telecom prices have fallen much more slowly than any other information technology. If this situation changed, it could completely alter the arguments here. But there is no obvious sign of that occurring.

## Acknowledgements

Many people have helped me gather this information and present the results. Gordon Bell, Charlie Catmull, Gerd Heber, George Spix, Alex Szalay, and Dan Worthheimer helped me characterize various computations. Ian Foster and Andrew Herbert helped me present the argument more clearly.